\documentclass[12pt]{iopart}

\usepackage{graphicx}
\begin{document}

\title[Strange Particle Production and Elliptic Flow from CERES]{Strange Particle Production and Elliptic Flow from CERES}

\author{J Milo\v{s}evi\'c (for the CERES Collaboration)}

\address{Physikalisches Institut der Universit\"{a}t Heidelberg, D 69120, Heidelberg, Germany}
\ead{jmilos@physi.uni-heidelberg.de}
\begin{abstract}
Elliptic flow measurements as a function of $p_{T}$ of charged ($\pi^{\pm}$
and low-$p_{T}$ protons) and strange ($\Lambda$ and $K_{S}^{0}$) particles
from Pb+Au collisions at 158~AGeV/c are presented, together with
measurements of $\phi$ and $K^{0}_{S}$ meson production. A mass ordering
effect was observed. Scaling to the number of constituent quarks and
transverse rapidity $y^{fs}_{T}$ scaling are presented. The results are
compared with results from the NA49 and STAR experiments and with
hydrodynamical calculations. For the first time in heavy-ion collisions,
$\phi$ mesons were reconstructed in the same experiment both in the
$K^{+}K^{-}$ and in the $e^{+}e^{-}$ decay channels. The obtained transverse
mass distributions of $\phi$ mesons are compared with results from the NA49
and NA50 experiments. The yield and the inverse slope parameter of the
$K_{S}^{0}$ mesons were reconstructed from two independent analyses. Our
results are compared with those from the NA49 and NA57 experiments.
\end{abstract}

\submitto{\JPG}
\maketitle

\section{Introduction}
Elliptic flow is described by the differential second Fourier coefficient of
the azimuthal momentum distribution $v_{2}({\cal D})=\langle \cos(2\phi)
\rangle_{\cal D}$ \cite{Ollit92,Bar94,PosVol}. The brackets denote averaging
over many particles and events, and ${\cal D}$ represents a phase-space window
in the $(p_{T},y)$ plane in which $v_{2}$ is calculated. The azimuthal angle
$\phi$ is measured with respect to the reaction plane defined by the impact
parameter vector $\vec b$ and the beam direction. For non-central collisions
($b \neq 0$), $v_{2}$ is an important observable due to its sensitivity to the
EoS, and through it to a possible phase transition to the QGP. Since we could
identify protons via $dE/dx$ only at low $p_{T}$, the $v_{2}$ of the $\Lambda$
is important because this is a baryon as well. In comparison to the elliptic
flow of pions and $K^{0}_{S}$ mesons the $\Lambda$ flow can be used to check
the mass ordering effect and for comparison to hydrodynamical
predictions. Testing the differential flow measurements of different particle
species against different scaling scenarios may yield additional information
about the origin of flow.

As strangeness enhancement has been suggested as a signature of the deconfined
stage \cite{Koch}, understanding of the $\phi$ and $K_{S}^{0}$ meson
production is important as here hidden and open strangeness are
involved. The study of $\phi$ yields in different decay channels is important
in light of a possible modification of the $\phi$ mass, width and the
branching ratios near the phase boundary.

\section{Experiment}

The CERES experiment consists of two radial Silicon Drift Detectors (SDD), two
Ring Imaging CHerenkov (RICH) detectors and a radial drift Time Projection
Chamber (TPC). The CERES spectrometer covers $\eta=2.05-2.70$ with full
azimuthal acceptance. The two SDDs are located at 10 and 13~cm downstream of a
segmented Au target. They were used for the tracking and vertex
reconstruction. The purpose of the RICH detectors is electron
identification. The new radial-drift TPC operated inside a magnetic field with
a maximal radial component of 0.5~T providing a precise determination of the
momentum. Charged particles emitted from the target are reconstructed by
matching track segments in the SDD and in the TPC using a momentum-dependent
matching window. A more detailed description of the CERES experiment can be
found in \cite{Mar}. For the flow analysis, we used 30$\cdot 10^{6}$ Pb+Au
events at 158~AGeV/c collected in the year of 2000 data taking period. Of
these, $91.2\%$ were triggered on $\sigma/\sigma_{geo} \le 7\%$, and $8.3\%$
events with $\sigma/\sigma_{geo} \le 20\%$. The $\phi$ meson analysis in the
kaon (dilepton) channel used 24$\cdot 10^{6}$ (18$\cdot 10^{6}$) events taken
with the most central trigger.

\section{Methods of strange particle reconstruction}
\label{Methods}

The $\Lambda$ particles were reconstructed via the decay channel $\Lambda
\rightarrow p+\pi^{-}$ with a $BR=63.9\%$ and $c\tau=7.89$~cm
\cite{PPB04}. Due to the late decay of the $\Lambda$ particle, as candidates
for $\Lambda$ daughters, only those TPC tracks which have no match to a SDD
track were chosen. Partial particle identification (PID) was performed using
$dE/dx$ information from the TPC by applying a $\pm 1.5 \sigma$ ($+1 \sigma$)
window around the momentum dependent Bethe-Bloch value for pions (protons). On
the pair level, a $p_{T}$ dependent opening angle cut is applied, in addition
to a cut in the Armenteros-Podalanski variables ($q_{T} \le 0.125$~GeV/c and
$0 \le \alpha \le 0.65$) to suppress $K_{S}^{0}$. With these cuts values for
$S/B\approx0.04$ and $S/\sqrt{B}\approx500$ were obtained \cite{Jovan}.

The $K_{S}^{0}$ particles were reconstructed via the decay channel
$K_{S}^{0}\rightarrow \pi^{+}+\pi^{-}$ with a $BR=68.95\%$ and
$c\tau=2.68$~cm \cite{PPB04}. Partial PID for $\pi^{+}$ and $\pi^{-}$
was performed by applying a $\pm 1.5 \sigma$ window around the momentum
dependent Bethe-Bloch energy loss value for pions. As the $K_{S}^{0}$ particle
comes from a primary vertex, a possibility to suppress fake track combinations
is given by a cut (0.02~cm) on the radial distance between the point where the
back extrapolated momentum vector of the $K_{S}^{0}$ candidate intersects the
$x-y$ plane and the primary vertex. In addition, a cut of 1~cm on the
z-position of the secondary vertex was applied. In this approach, the values
of $S/B \approx 0.92$ and $S/\sqrt{B} \approx 500$ were obtained
\cite{Wilrid,Jovan}.

In order to remove the effect of autocorrelations, tracks which were chosen as
candidates for daughter particles were not used for the determination of the
reaction plane orientation. In the case of $\Lambda$ particle reconstruction,
the combinatorial background was determined by ten random rotations of
positive daughter tracks around the beam axis and constructing the invariant
mass distribution, while in the case of $K^{0}_{S}$ particle reconstruction,
the mixed event technique was used.

$\Lambda$ ($K^{0}_{S}$) particles were reconstructed in $y$-$p_{T}$-$\phi$
bins. We used the area under the peak, obtained by fitting the invariant mass
distribution with a Gaussian, to measure the yield of $\Lambda$ ($K^{0}_{S}$)
in a given bin. Plotting the yield versus $\phi$ for different $p_{T}$ and $y$
values one can construct the $dN_{\Lambda(K^{0}_{S})}/d\phi$
distribution. Fitting these distributions with a function
$c[1+2v_{2}'\cos(2\phi)]$, it is possible to extract the observed differential
$v_{2}'$ values. The obtained $v_{2}'$ coefficients were corrected for the
reaction plane resolution via $v_{2}=v_{2}'/\sqrt{2\langle
  \cos[2(\Phi_{a}-\Phi_{b})]\rangle}$ \cite{PosVol}. Here, $\Phi_{a}$ and
$\Phi_{b}$ denote the azimuthal orientations of reaction planes reconstructed
from two random subevents. In the case of the $\pi^{\pm}$ elliptic flow
analysis, subevents are formed from positive and negative pions
separately. Using the method of subevents, correction factors were calculated
for different centrality bins. In all 3 analyses ($\Lambda$, $K^{0}_{S}$ and
$\pi^{\pm}$) similar values were obtained. The corresponding resolution ranges
from about 0.16 to 0.31, depending on the centrality.

Due to the small statistics of strange particles, the differential elliptic
flow analysis was performed for only two centrality classes. The huge
statistics of $\pi^{\pm}$ allowed to perform the differential elliptic flow
analysis in six centrality bins. As we used the combination of data taken with
different triggers, the centrality is characterized by a weighted mean
centrality $\langle \frac{\sigma}{\sigma_{geo}} \rangle$ calculated using
the numbers of TPC tracks as statistical weights \cite{Jovan}.

\section{Results}

\begin{figure}[h]
\begin{minipage}[c]{.33\textwidth}
\includegraphics[height=5.2cm]{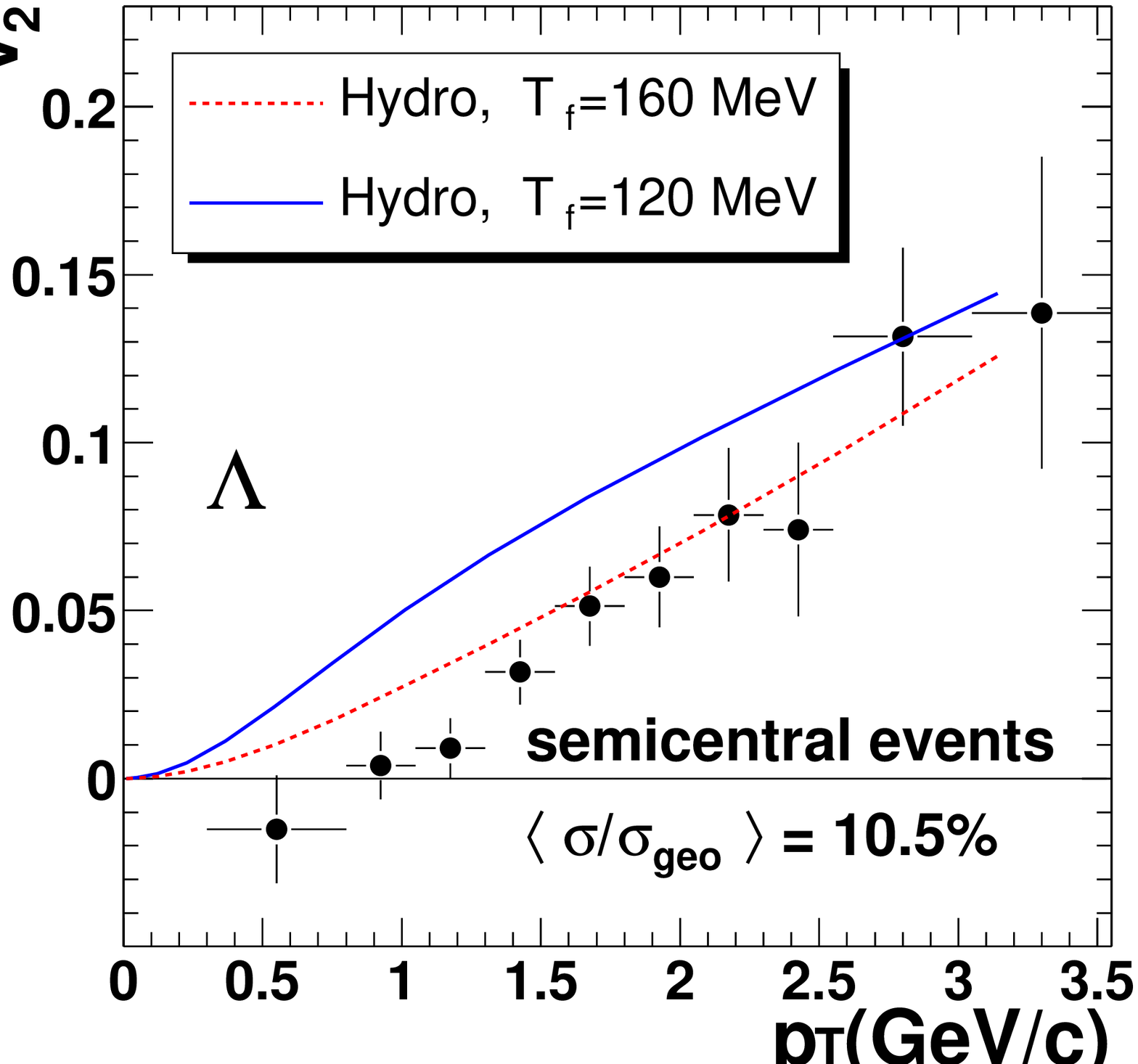}
\end{minipage}
\begin{minipage}[c]{.33\textwidth}
\includegraphics[height=5.2cm]{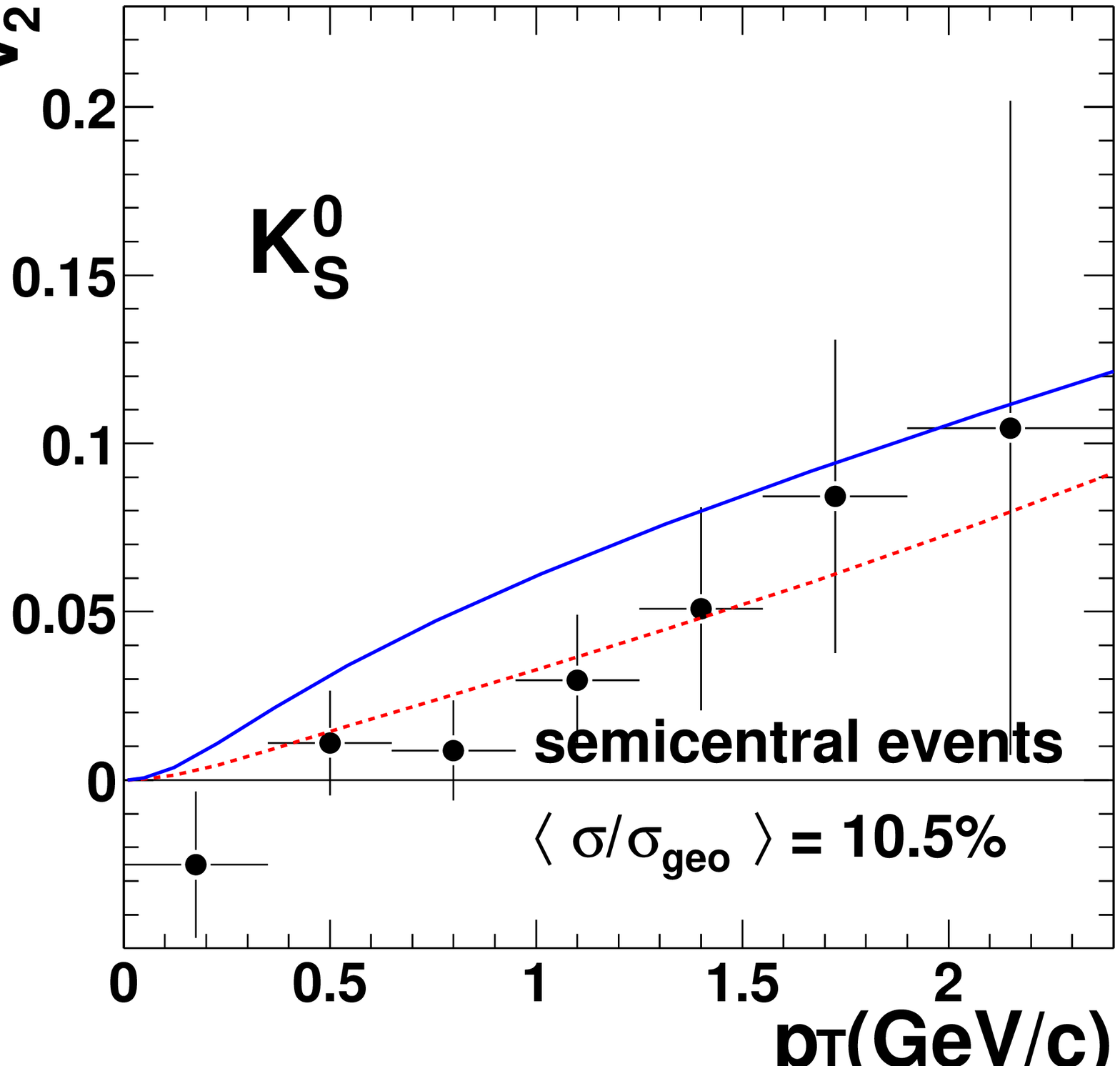}
\end{minipage}
\begin{minipage}[c]{.33\textwidth}
\includegraphics[height=5.2cm]{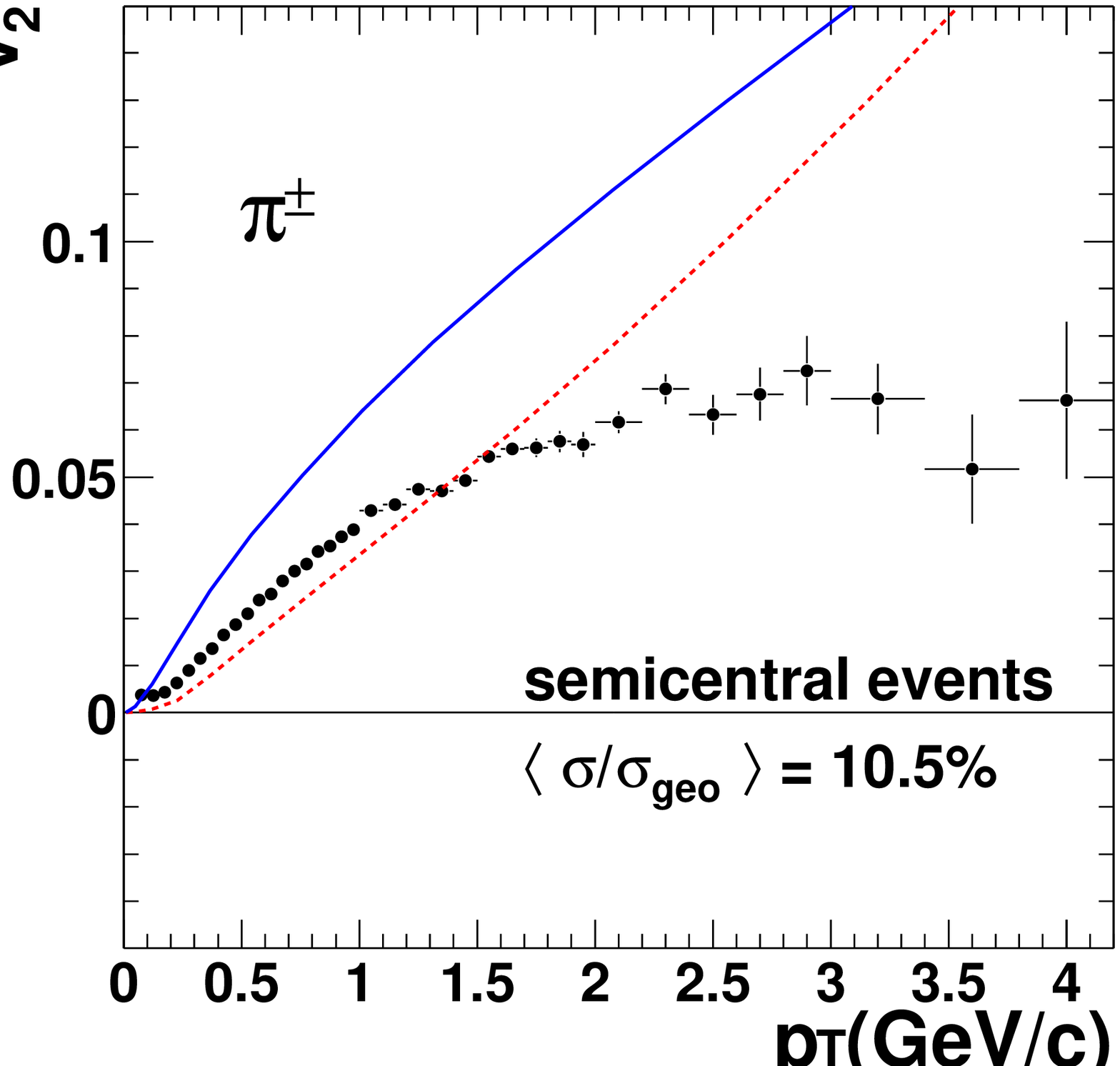}
\end{minipage}
  \caption{The $\Lambda$ (left), $K^{0}_{S}$ (middle) and $\pi^{\pm}$ (right)
    elliptic flow $\it vs$ transverse momentum in semicentral
    events. Hydrodynamical predictions are presented for two freeze-out
    temperatures: $T_{f}=120$~MeV (solid) and $T_{f}=160$~MeV
    (dotted). \label{fig:hydro}}
\end{figure}

In Fig.~\ref{fig:hydro} are shown the resulting $p_{T}$ dependences of
$v_{2}$ for three particle species. An increase of the elliptic flow
magnitude {\it vs} $p_{T}$ for all three particle species is visible. In the
case of $\Lambda$ elliptic flow, the absolute systematic error $\Delta v_2$,
estimated from two different ways of $\Lambda$ reconstruction, is $+0.001
\atop -0.007$ for $p_{T} < 1.6$~GeV/c and $+0.00 \atop -0.02$ for $p_{T} >
1.6$~GeV/c which is small compared to the statistical errors. Particles are
accepted as $\pi^{\pm}$ if their TPC $dE/dx$ is within a $\pm 1.5\sigma$
window around the nominal Bethe-Bloch value for pions. The HBT contribution to
the $\pi^{\pm}$ elliptic flow is subtracted using the procedure described in
\cite{Dinh99}. Separately calculated elliptic flow of $\pi^{+}$ and $\pi^{-}$
shows that the averaged difference between them is $\approx 0.003$ in both
$\eta$ and $y$, which can be attributed to the contamination of protons in
$\pi^{+}$ sample. Comparing results obtained from two independent analysis
methods we concluded that the overall absolute systematic error in $\pi^{\pm}$
elliptic flow measurements is not bigger than 0.0036.

The elliptic flow results are compared with the hydrodynamical calculations
done by P. Huovinen based on \cite{Kolb01,Pasi05}. The calculation was done in
2+1 dimensions with initial conditions fixed via a fit to the $p_{T}$ spectra
of negatively charged particles and protons in Pb+Pb collisions at 158~A GeV/c
\cite{Kolb99}. The underlying EoS assumes a first order phase transition to a
QGP at a critical temperature of $T_{c}=165$~MeV. The hydrodynamical
predictions were calculated with 2 freeze-out temperatures, $T_{f}=120$~MeV
and $T_{f}=160$~MeV. The model prediction with the lower freeze-out
temperature of $T_{f}=120$~MeV overpredicts the data, while rather good
agreement can be achieved with a higher freeze-out temperature of
$T_{f}=160$~MeV (this is however not the preferred value considering the
proton $p_{T}$ spectra).

\begin{figure}[h]
\begin{minipage}[c]{.33\textwidth}
\includegraphics[height=5.2cm]{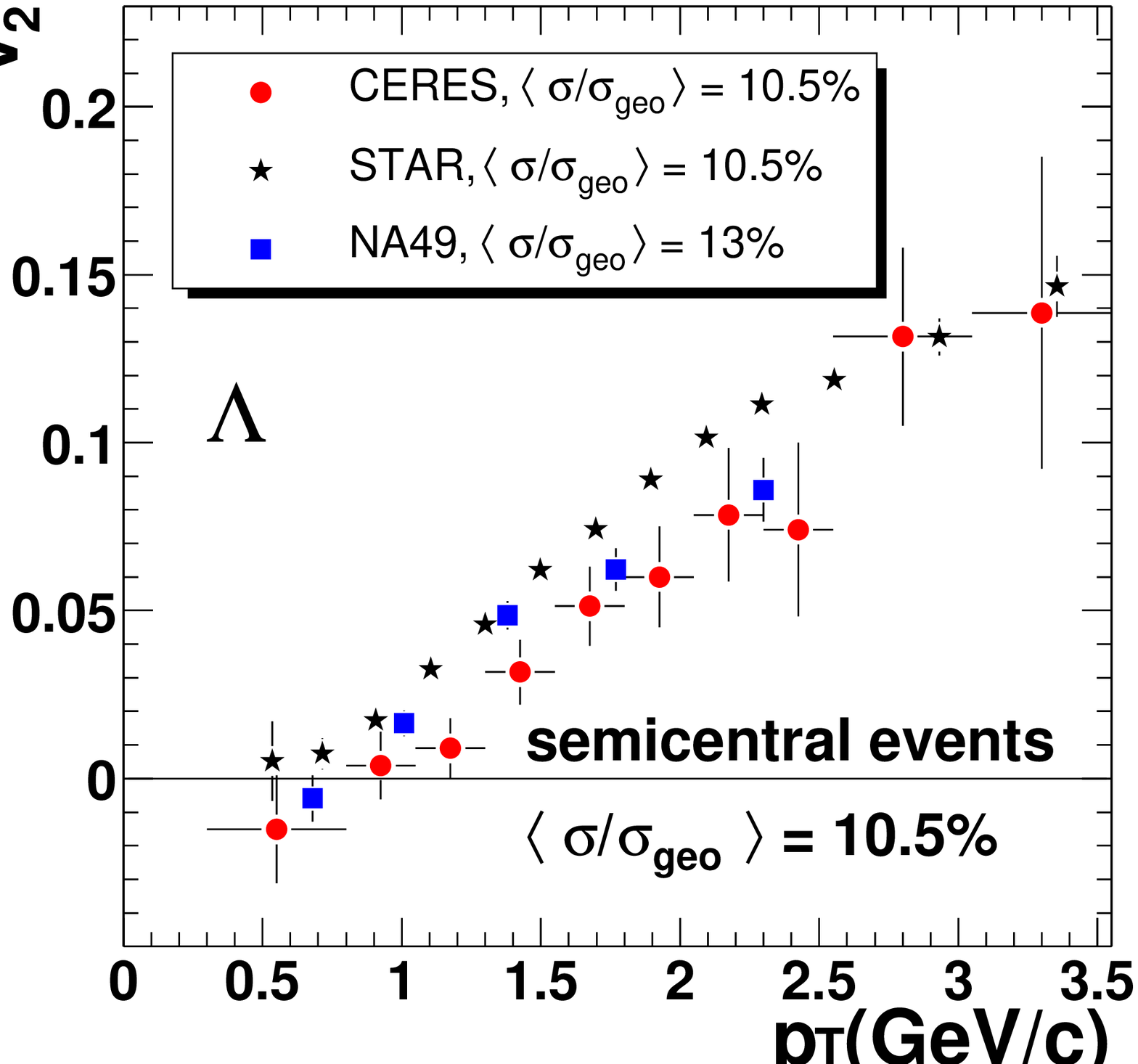}
\end{minipage}
\begin{minipage}[c]{.33\textwidth}
\includegraphics[height=5.2cm]{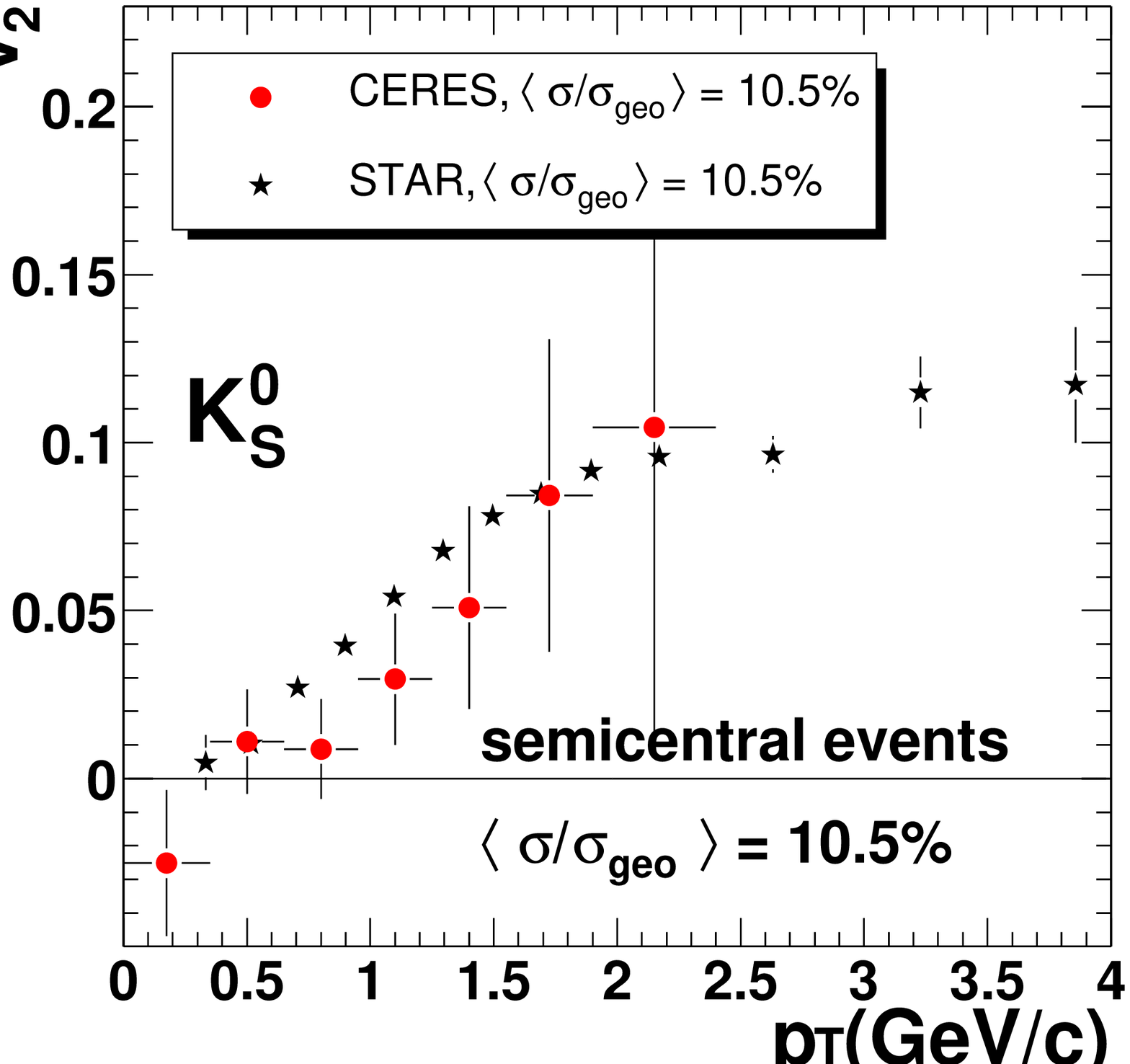}
\end{minipage}
\begin{minipage}[c]{.33\textwidth}
\includegraphics[height=5.2cm]{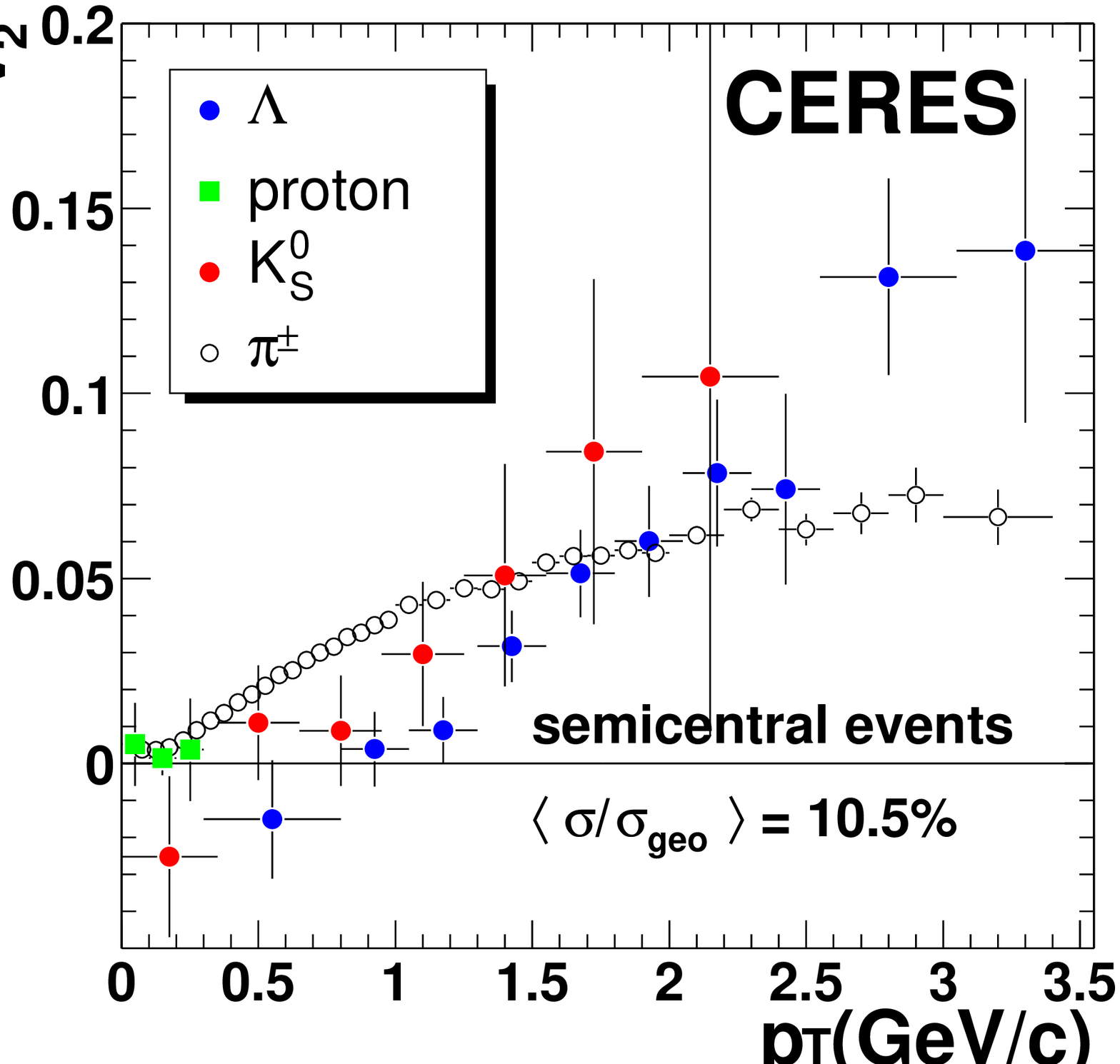}
\end{minipage}
  \caption{Comparison of $\Lambda$ (left) and $K^{0}_{S}$ (middle) elliptic
    flow measured by CERES, STAR and NA49. Comparison between the elliptic
    flow magnitude of the $\pi^{\pm}$, low-$p_{T}$ protons, $\Lambda$, and
    $K_{S}^{0}$ in semicentral events (right). \label{fig:compare_RHIC_NA49}}
\end{figure}

A comparison of the CERES data to results from NA49 \cite{Stef05} at the same
energy ($\sqrt{s_{NN}}=17$~GeV) and to STAR results \cite{Oldenburg05} at
$\sqrt{s_{NN}}=200$~GeV is shown in Fig.~\ref{fig:compare_RHIC_NA49}. The NA49
and CERES data are in very good agreement. After rescaling the STAR results
to the centrality used in the CERES experiment, the $v_{2}$ values measured at
RHIC are $15-20\%$ higher due to the higher beam energy. In
Fig.~\ref{fig:compare_RHIC_NA49} (right), the elliptic flow magnitude of the
$\pi^{\pm}$, $K_{S}^{0}$, low momentum protons, and $\Lambda$ measured by
CERES are compared. A mass ordering effect is observed. At small $p_{T}$, up
to $\approx1.5$~GeV/c, $v_{2}(\Lambda)<v_{2}(K_{S}^{0})<v_{2}(\pi^{\pm})$. In
the region of high $p_{T}$, above $\approx2$~GeV/c, the tendency is the
opposite. As proton and $\Lambda$ hyperon have similar masses and 3 valence
quarks each, the $v_{2}$ of low momentum identified protons is considered as a
natural continuation of $\Lambda$ $v_{2}(p_{T})$ dependence in the region of
small $p_{T}$. The indication of a possible undershoot to negative values is
tantalizing but not significant in view of the statistical errors.

 \begin{figure}[h]
   \begin{minipage}[h]{.48\textwidth}
     \includegraphics[height=5.2cm] {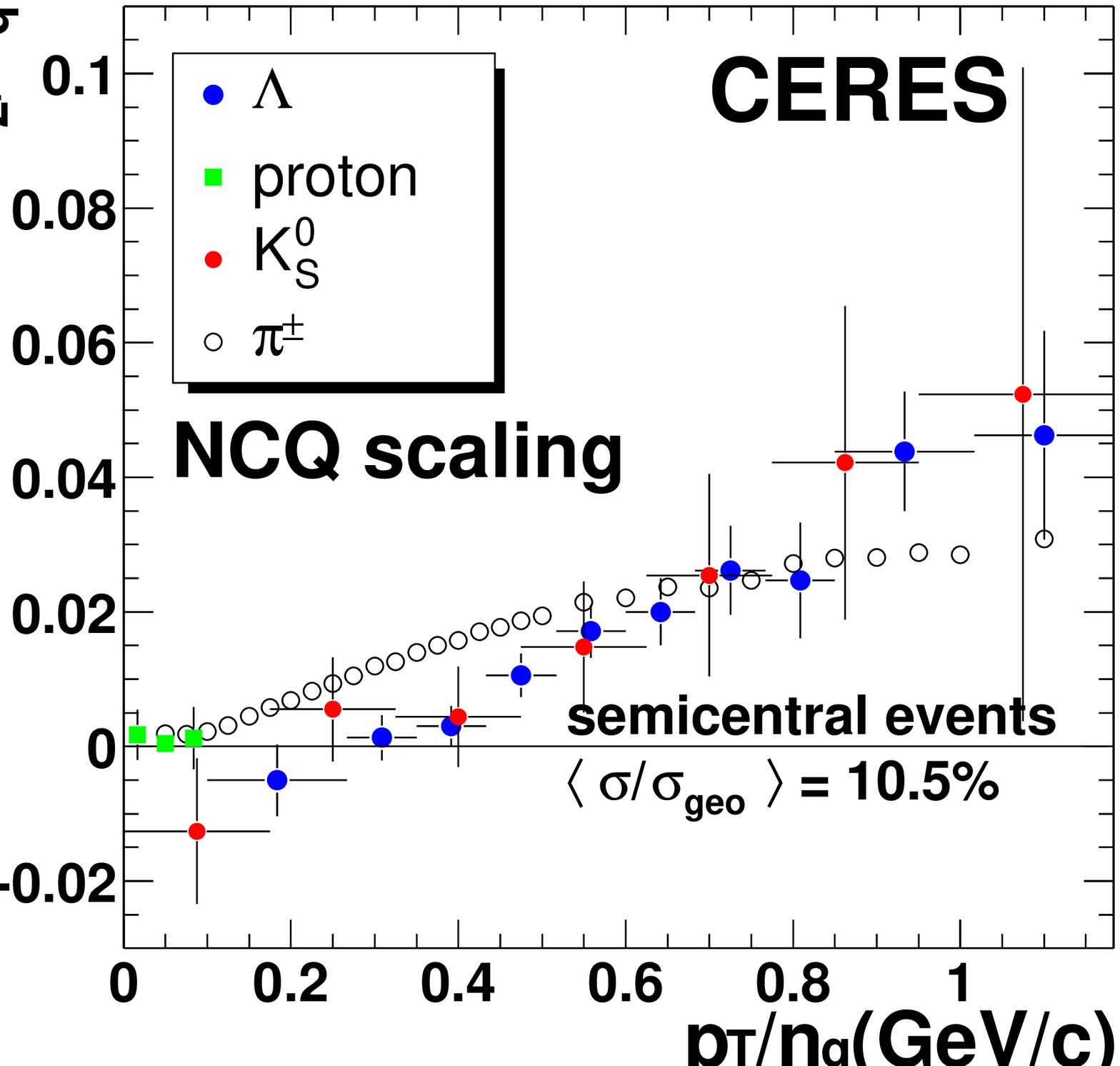}
   \end{minipage}
   \begin{minipage}[h]{.48\textwidth}
     \includegraphics[height=5.2cm] {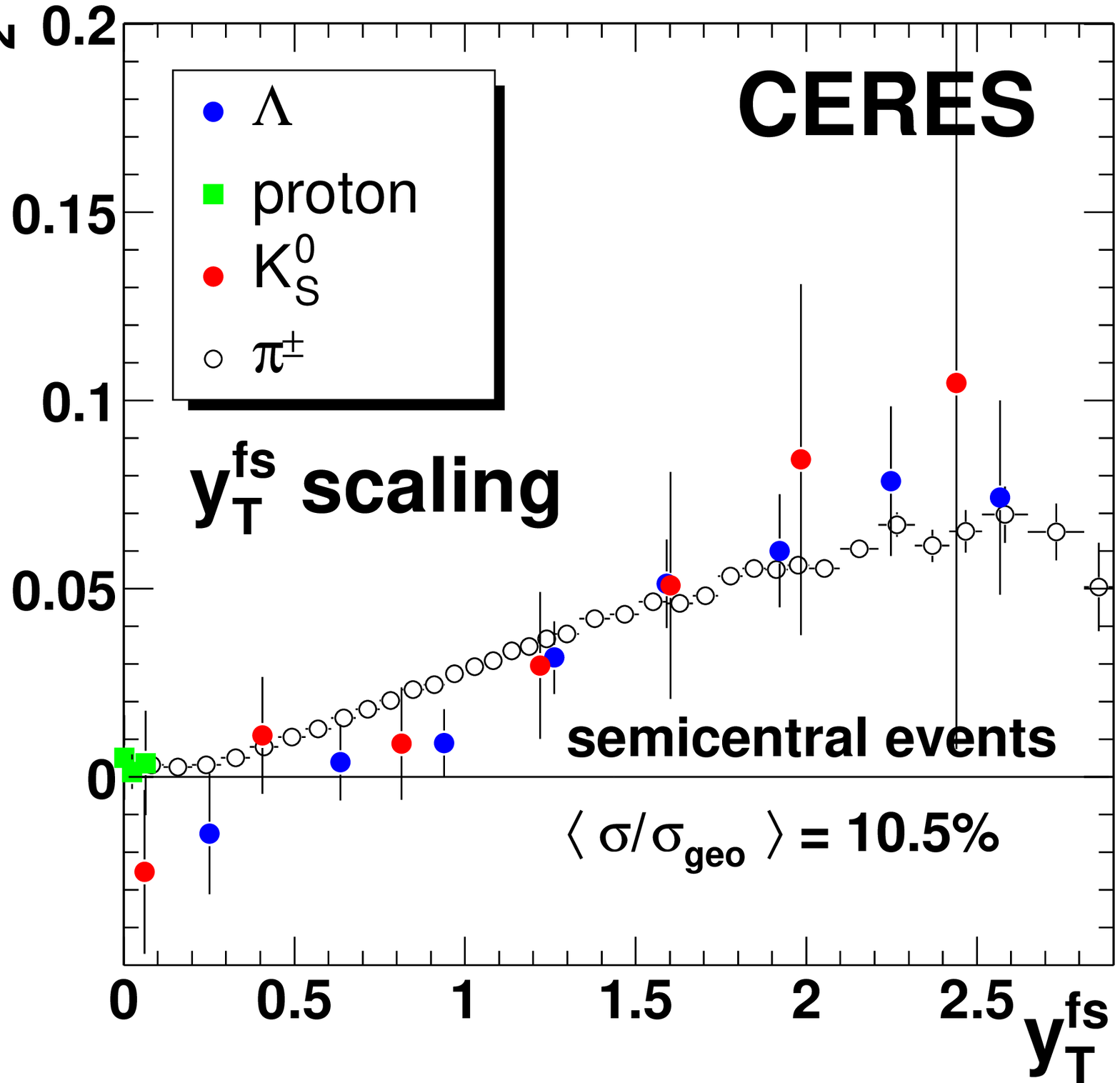}
   \end{minipage}
   \caption{Comparison between the elliptic flow magnitude of $\pi^{\pm}$,
     low-$p_{T}$ protons, $\Lambda$, and $K_{S}^{0}$ scaled to the number
     of the constituent quarks (left) and to the $y^{fs}_{T}$ variable
     (right). \label{fig:scaled_all_new}}
 \end{figure}

Fig.~\ref{fig:scaled_all_new} (left) shows the scaled elliptic flow magnitude
$v_{2}/n_{q}$ for $\pi^{\pm}$, $K_{S}^{0}$, low-$p_{T}$ protons and $\Lambda$ 
plotted against $p_{T}/n_{q}$ in semicentral events. Here, $n_{q}$ denotes the
number of the constituent quarks. There is an indication
that high $p_{T}$ particles ($p_{T}>1.5$~GeV/c) show scaling behavior. A
similar behavior is observed by the STAR experiment at RHIC
\cite{Oldenburg05}. This is consistent with the coalescence mechanism where
co-moving quarks with high $p_{T}$ form hadrons. In this case scaling to the
number of the constituent quarks shows the original momentum space azimuthal
anisotropy formed at the early stage of the collision.

Within the Buda-Lund model of hydrodynamics \cite{Cso96}, a scaling of elliptic
flow of different particle species has been suggested \cite{Cso03,Csa04} when
instead transverse momentum the transverse rapidity is used. We use their
scaling variable $y^{fs}_{T}$ \cite{Taran05} and show, in
Fig.~\ref{fig:scaled_all_new} (right), the results for $\pi^{\pm}$,
$K_{S}^{0}$, low-$p_{T}$ protons and $\Lambda$ in semicentral
events. Within statistical errors a reasonable scaling is observed for all
particles. This may indicate a hydrodynamic behavior of matter created in
central heavy-ion collisions at the highest SPS energy.

 \begin{figure}[h]
   \begin{minipage}[h]{.49\textwidth}
     \includegraphics[height=5.5cm] {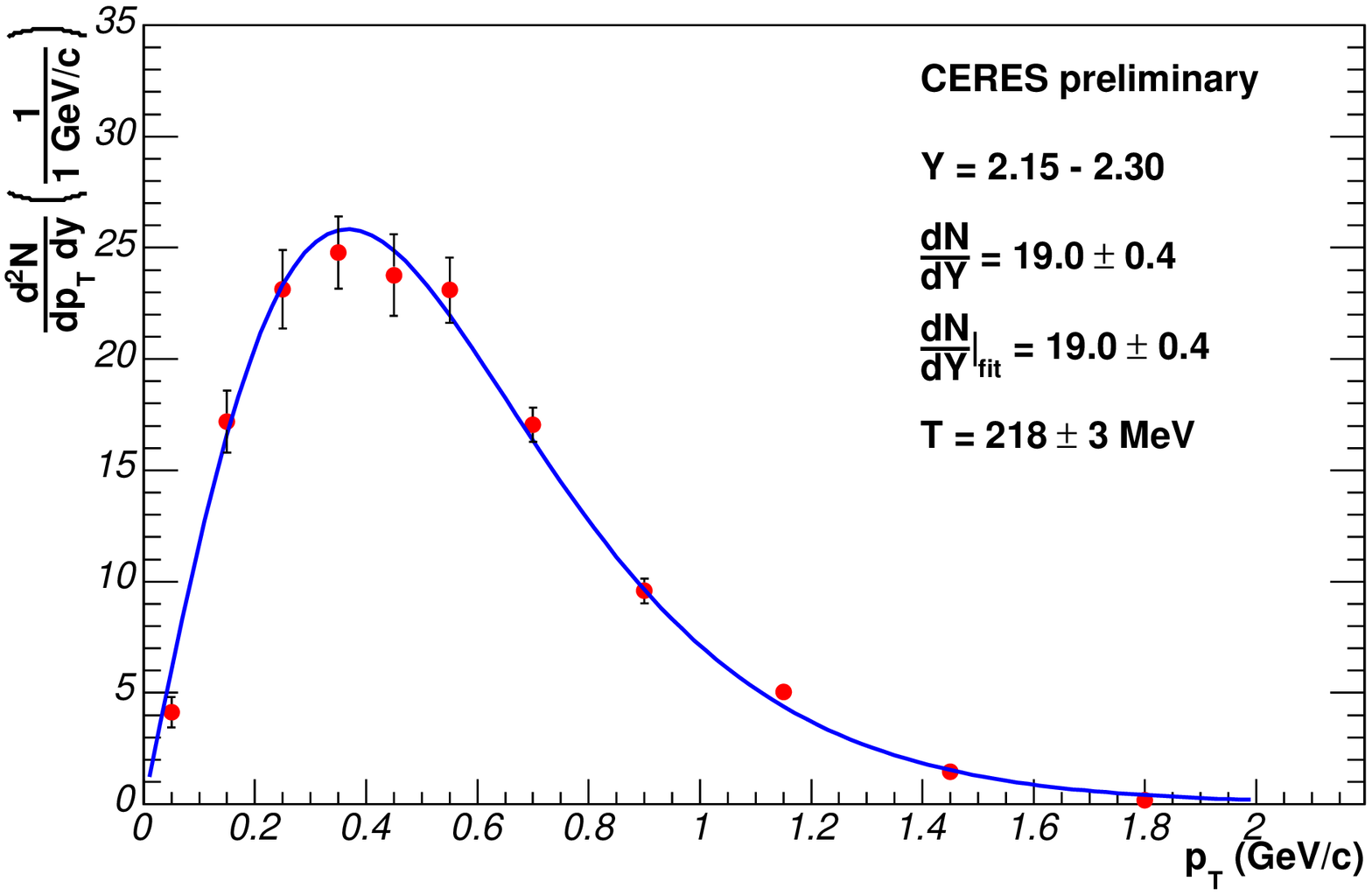}
   \end{minipage}
   \begin{minipage}[h]{.49\textwidth}
     \includegraphics[height=5.5cm] {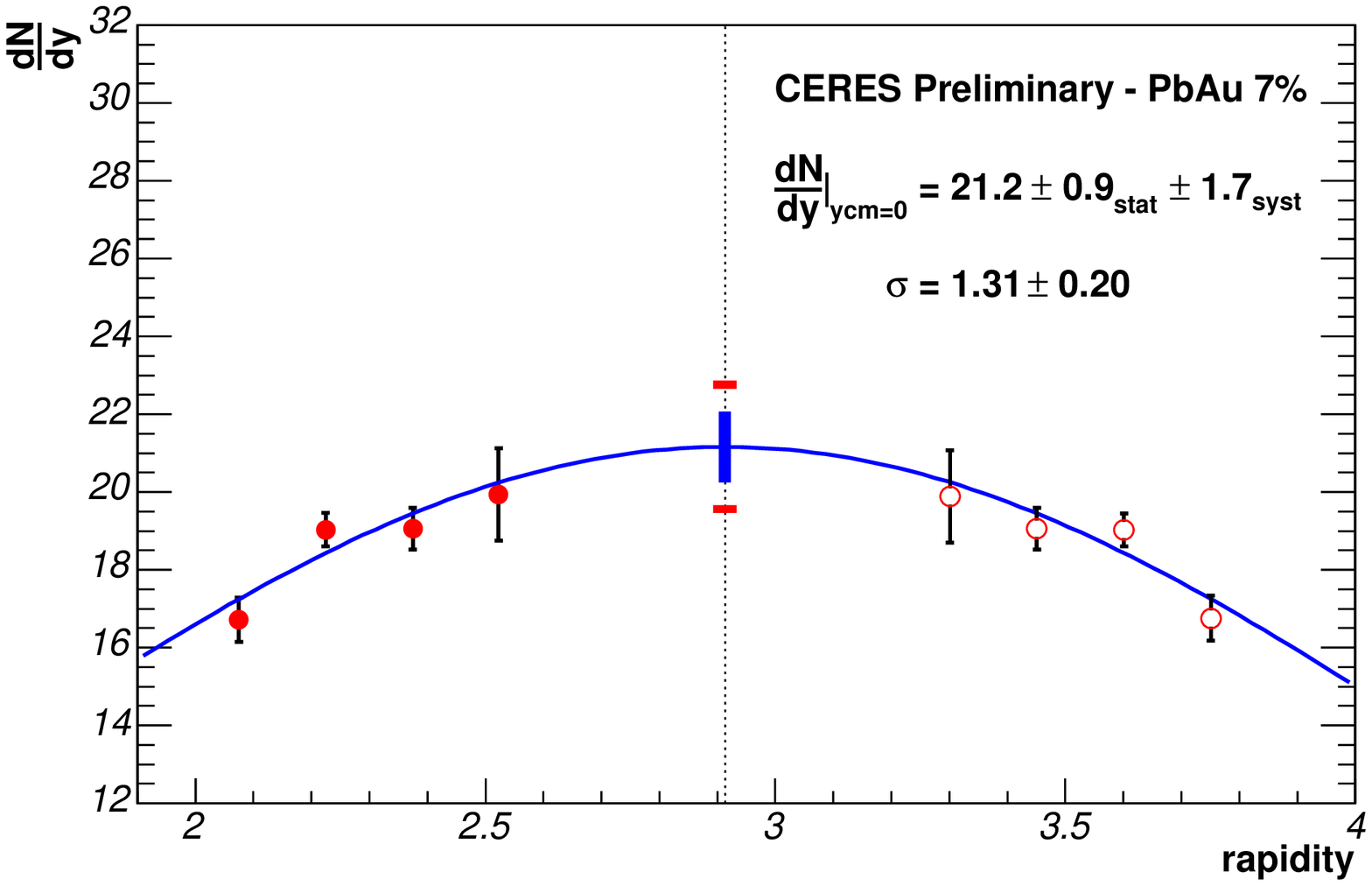}
   \end{minipage}
   \caption{Transverse momentum and rapidity $K_{S}^{0}$ spectra from the
     $K^{0}_{S}$ analysis performed without PID and without secondary vertex
     reconstruction \cite{Radom}. \label{fig:kaons}}
\end{figure}

\begin{figure}[h]
   \begin{minipage}[h]{.49\textwidth}
     \includegraphics[height=6.0cm] {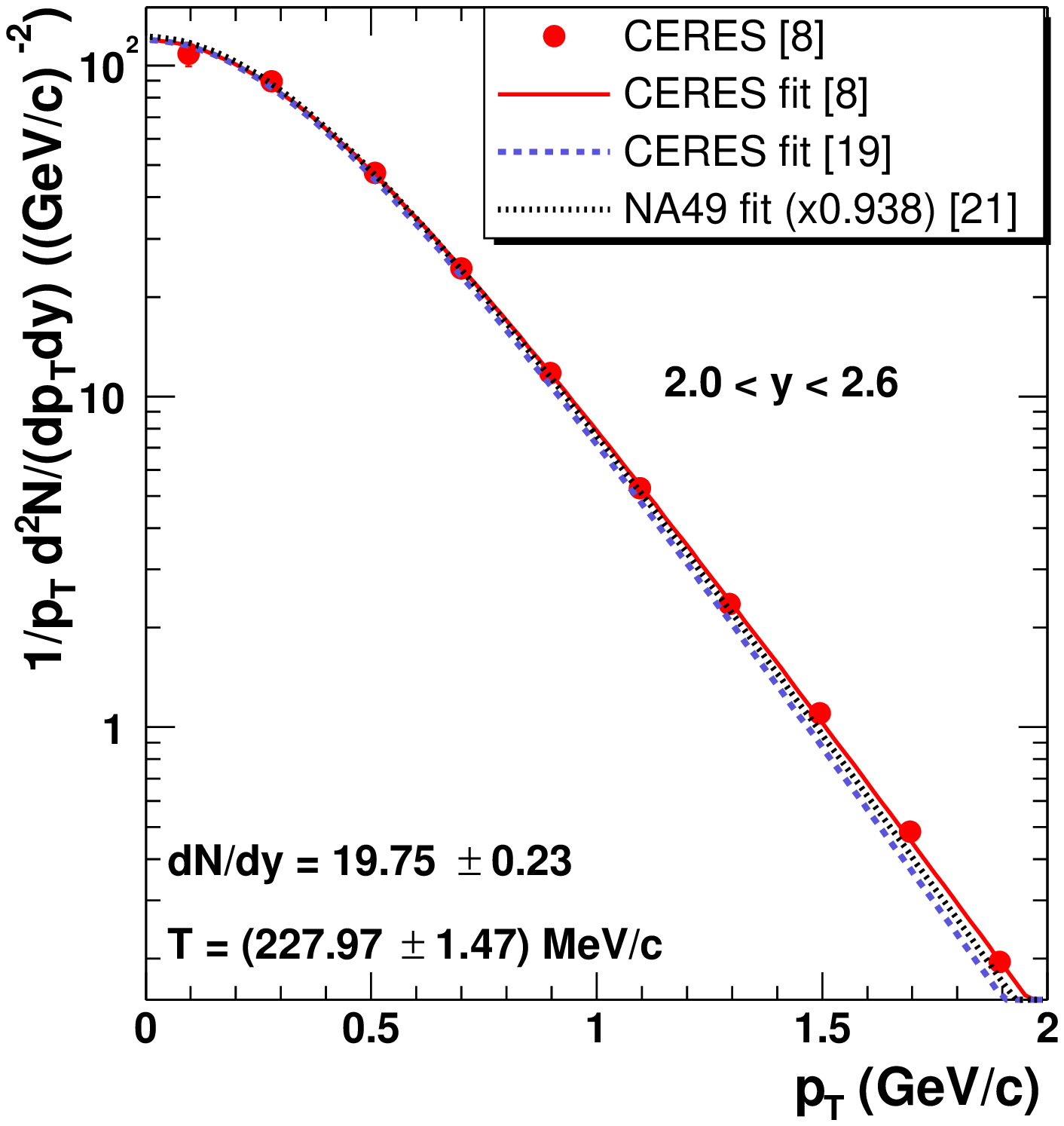}
   \end{minipage}
   \begin{minipage}[h]{.49\textwidth}
     \includegraphics[height=6.0cm] {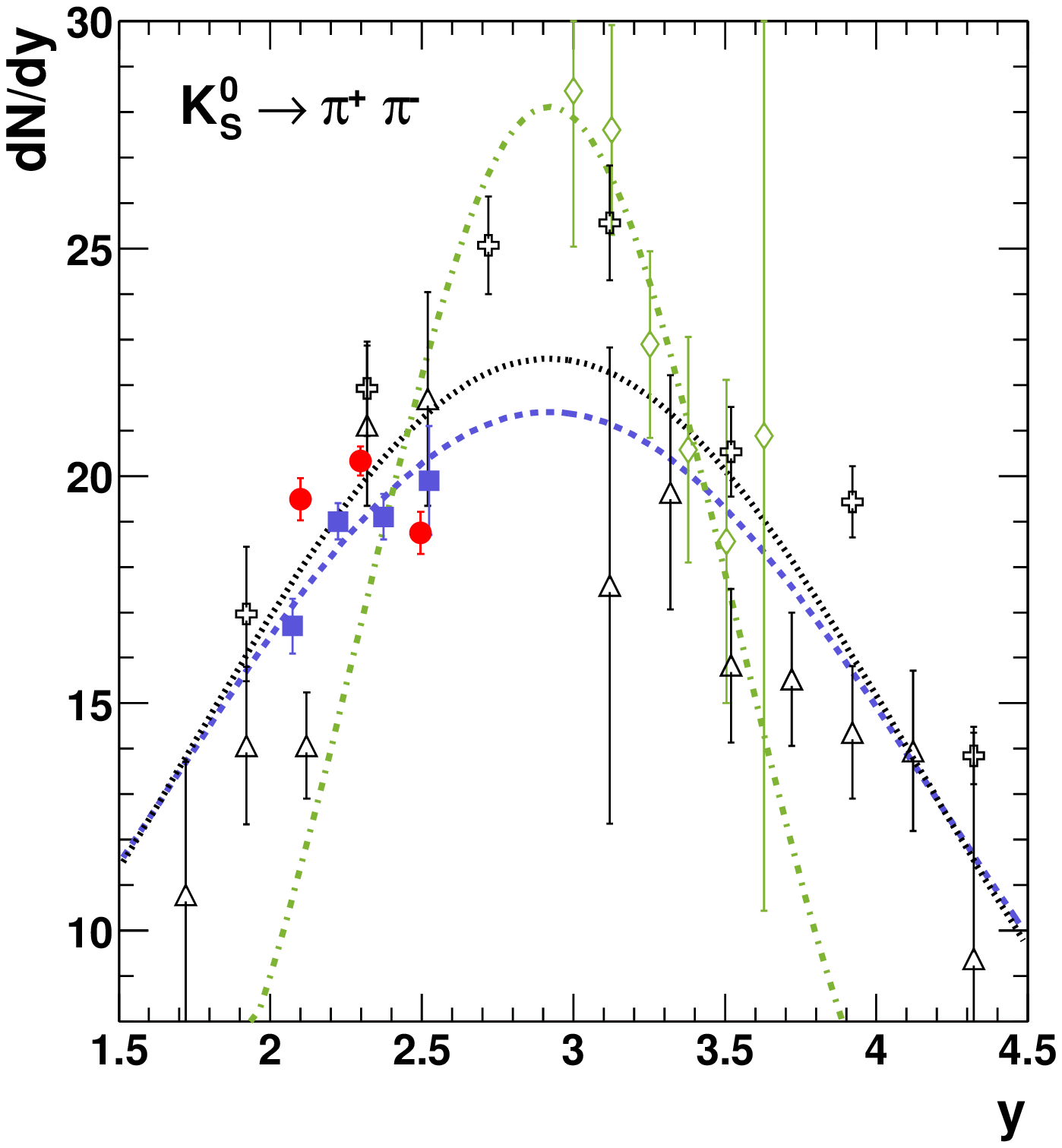}
   \end{minipage}
   \caption{Left: Transverse momentum $K_{S}^{0}$ spectrum from the
     $K^{0}_{S}$ analysis performed with secondary vertex
     reconstruction \cite{Wilrid}. Right: A comparison between CERES results
     (red circles \cite{Wilrid} and blue squares \cite{Radom}), published
     (open triangles) \cite{NA49} and preliminary (open crosses)
     \cite{NA49_nuclexp} NA49 results and NA57 data (green diamonds)
     \cite{NA57}. The black dotted line represents a fit to the charged kaon
     yield measured by the NA49, while blue (green) dotted line corresponds
     to a fit to the $K_{S}^{0}$ yield measured by the CERES
     (NA57). \label{fig:Ludolphs}} 
\end{figure}

Two independent analyses of the $K_{S}^{0}$ spectra were done using the CERES
data \cite{Radom,Wilrid}. The first one, performed without PID and without
secondary vertex reconstruction, is based on TPC information only
\cite{Radom}. A cut in the Armenteros-Podalanski plane was used in order to
suppress $\Lambda$ contamination. The $p_{T}$ and $y$ spectra are shown in
Fig.~\ref{fig:kaons}. An alternative approach of the $K_{S}^{0}$
reconstruction was performed without PID but with secondary vertex
reconstruction \cite{Wilrid} which is already described in
Section~\ref{Methods}. In both analyses, Pb+Au events taken with the most
central trigger were used. The $K_{S}^{0}$ transverse momenta spectrum
obtained with this analysis \cite{Wilrid} is shown in Fig.~\ref{fig:Ludolphs}
(left). The invariant multiplicity was fitted with an exponential fall-off
with transverse mass $m_{t}$. The yields and the inverse slope parameter $T$
of the $p_{T}$ spectra from the two analyses are in good agreement. A
comparison with results from other experiments is shown in
Fig.~\ref{fig:Ludolphs} (right). In order to match the centrality of the CERES
experiment, results
from the NA49 \cite{NA49,NA49_nuclexp} and NA57 \cite{NA57} experiments are
slightly rescaled. A rather good agreement between the NA49 analysis of
charged kaons and the CERES $K^{0}_{S}$ results in shape and yield was
found. The difference in the yield is only 5\%. The rapidity distribution of
$K^{0}_{S}$ observed by NA49 shows a similar shape as the one from CERES
(represented with the blue dotted line in Fig.~\ref{fig:Ludolphs} (right))
and a relatively good agreement in the yield. Within the CERES acceptance the
results agree with the NA57 data, although the NA57 fit does not.

The CERES experiment enabled for the first time at SPS to study simultaneously
the leptonic and charged kaon decay modes of the $\phi$ meson, which may shed
light onto the $\phi$ puzzle \cite{PRLMarin}. In order to obtain the $p_{T}$
spectrum of $\phi$ mesons, the invariant
\begin{figure}[h]
   \begin{minipage}[h]{.48\textwidth}
     \includegraphics[height=5.6cm] {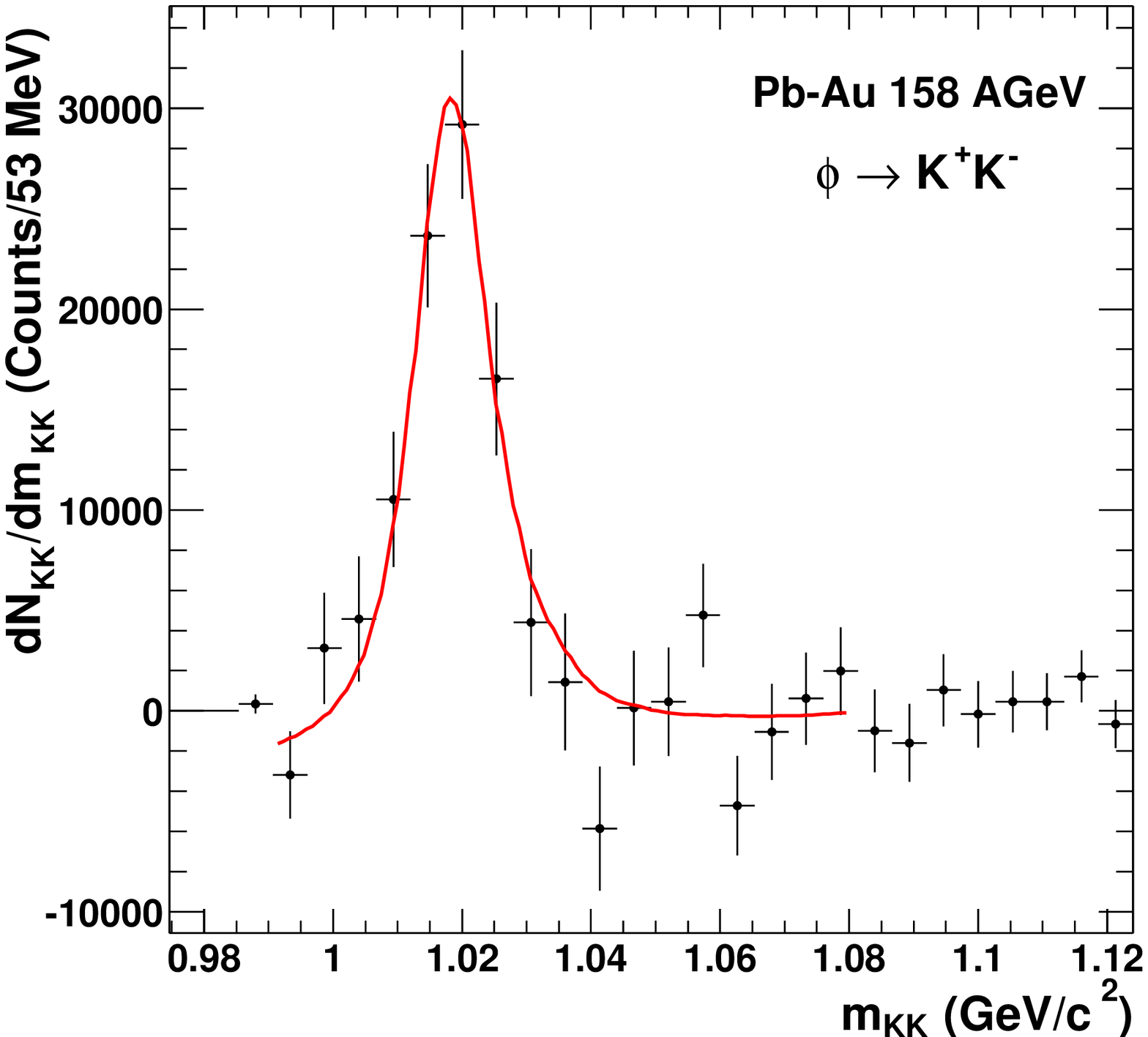}
   \end{minipage}
   \begin{minipage}[h]{.48\textwidth}
     \includegraphics[height=5.6cm] {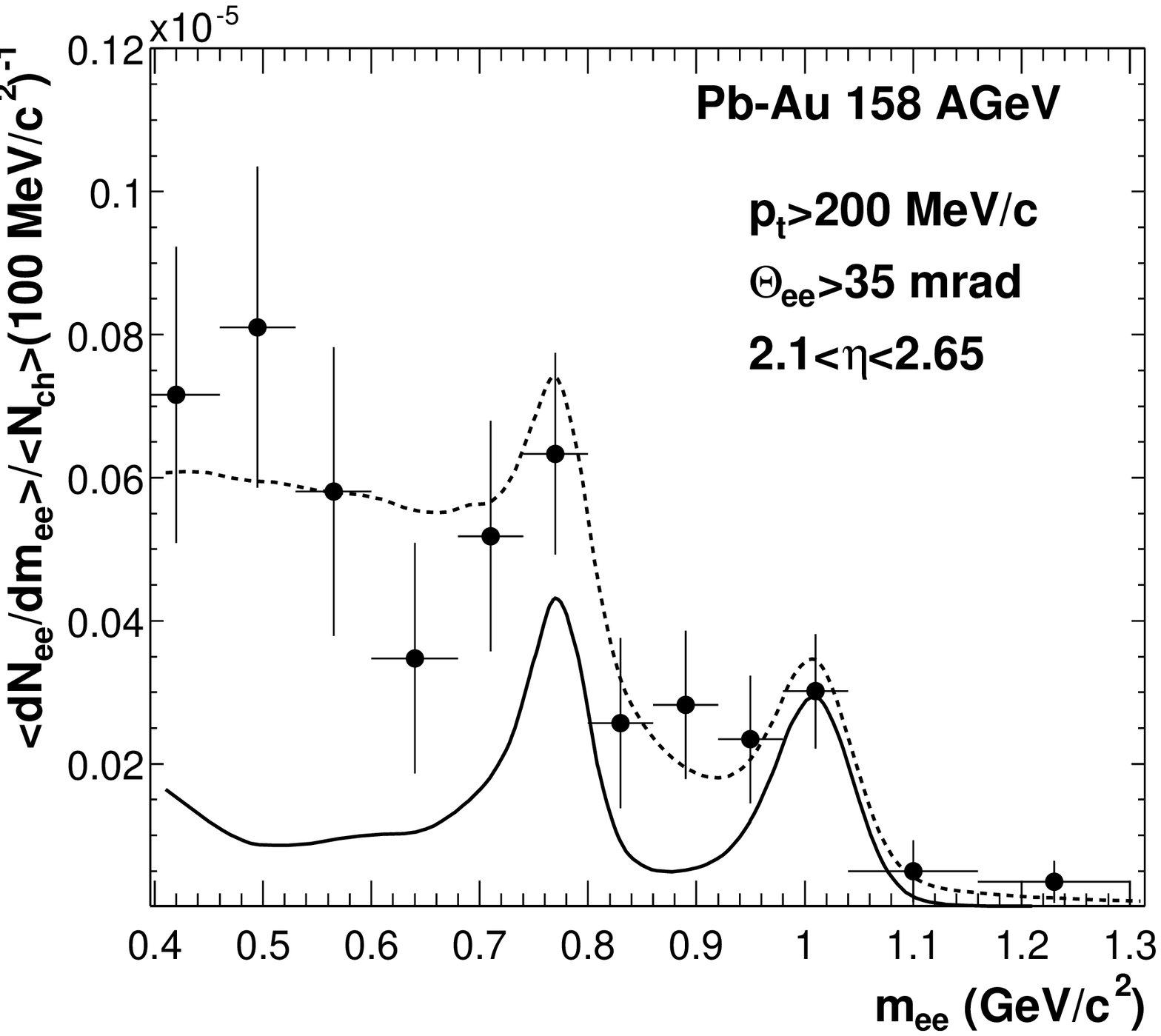}
   \end{minipage}
   \caption{Left: $K^{+}K^{-}$ invariant mass spectrum after background
     subtraction in $1.5$~GeV/c $<p^{\phi}_{T}<1.75$~GeV/c and
     $2.2<y^{\phi}<2.4$. Right: $e^{+}e^{-}$ invariant mass spectrum compared
     to the hadron decay cocktail (solid line) and to a model calculation
     assuming the dilepton yield form the QGP phase and an in medium spread
     $\rho$ (dashed line). \label{fig:ceres}} 
 \end{figure}
mass distributions of $K^{+}K^{-}$ pairs were constructed. The corresponding
distributions of the combinatorial background were calculated using the
mixed-event technique. An example is shown in Fig.~\ref{fig:ceres}
(left). To study $\phi$ mesons in the dilepton ($e^{+}e^{-}$) decay mode,
electrons are identified using the RICH detectors and the TPC $dE/dx$. The
main difficulties of reconstructing the $\phi$ meson in the dilepton channel
are the low branching ratio and huge combinatorial background. Details of how
to reduce the combinatorial background are explained in
\cite{Mar,Misko,Sergej}. The $e^{+}e^{-}$ invariant-mass spectrum, corrected
for the efficiency and normalized to the number of charged particles in the
acceptance is shown in Fig.~\ref{fig:ceres} (right). In the same figure are
shown the expectations from the hadron decay cocktail \cite{Sako}, as well as  
a model calculation where the cocktail $\rho$ contribution is replaced by an
explicit in-medium modification combined with continuous $\pi\pi$
annihilation \cite{Rap}. The later accounts very well for the data. The
inverse slope parameter of $T=$273$\pm$9(stat)$\pm$10(syst)~MeV and a rapidity
density $dN/dy$ of 2.05$\pm$0.14(stat)$\pm$0.25(syst) in the $K^{+}K^{-}$ mode
and $T=$306$\pm$82(stat)$\pm$40(syst)~MeV and
$dN/dy=$2.04$\pm$0.49(stat)$\pm$0.32(syst) in the dilepton mode are in good
agreement within errors. The data do not support a possible enhancement of the
$\phi$ yield in the dilepton over the hadronic channel by a factor larger than
1.6 at the 95\% CL.

 \begin{figure}[h]
   \begin{minipage}[h]{.48\textwidth}
     \includegraphics[height=5.6cm] {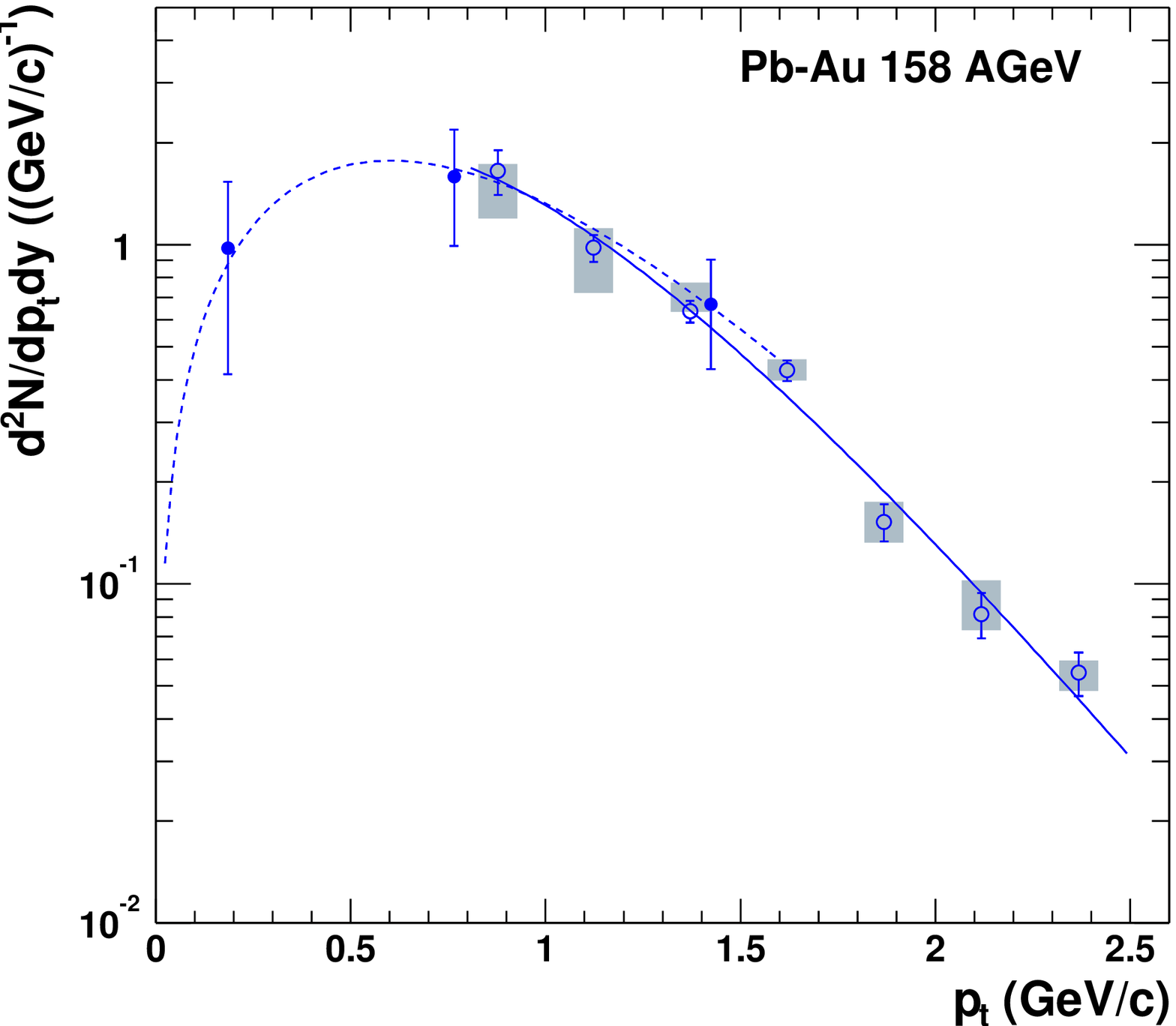}
   \end{minipage}
   \begin{minipage}[h]{.48\textwidth}
     \includegraphics[height=5.6cm] {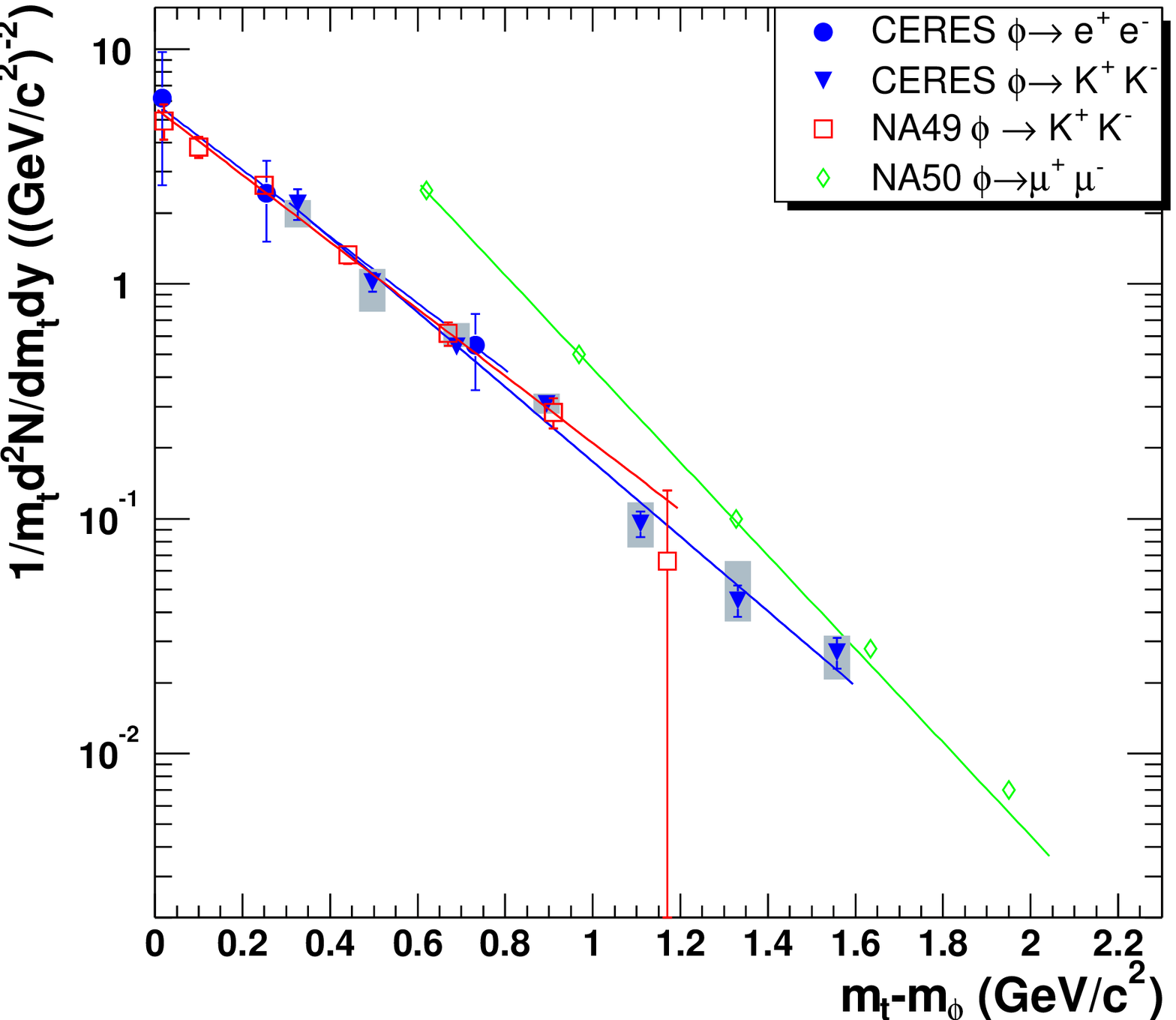}
   \end{minipage}
   \caption{Left: Acceptance and efficiency corrected $p_{T}$
     spectrum of $\phi$  measured in the $K^{+}K^{-}$ (open cycles)
     and $e^{+}e^{-}$ (closed circles) decay modes. Right: Scaled $m_{T}$
     distribution of $\phi$ mesons reconstructed in the $K^{+}K^{-}$
     (triangles) and $e^{+}e^{-}$ (circles) decay channels compared to the
     results from NA49 (squares) and NA50 (diamonds). \label{fig:marin}}
 \end{figure}

The $p_{T}$ dependence of the $\phi$ meson yield  measured in the $K^{+}K^{-}$
and $e^{+}e^{-}$ decay channels, corrected for the acceptance and efficiency,
is shown in Fig.~\ref{fig:marin} (left). The results are in very good
agreement. After accounting for the slightly different measurement conditions,
a comparison between CERES results and the existing Pb+Pb systematics
\cite{Ror} is shown in Fig.~\ref{fig:marin} (right). The CERES results are in
good agreement with the results from NA49 measured in the kaon channel. On
the other hand, CERES data in the $K^{+}K^{-}$ channel do not agree with NA50
results.


\section*{References}
\begin{thebibliography}{10}
\bibitem{Ollit92} Ollitrault J-Y 1992 {\it Phys. Rev.} D {\bf 46} 229
\bibitem{Bar94} Barrette J {\it et al} [E877 Collaboration] 1994
  {\it Phys. Rev. Lett.} {\bf 73} 2532
\bibitem{PosVol} Poskanzer A M and Voloshin S A 1998
  {\it Phys. Rev.} C {\bf 58} 1671
\bibitem{Koch} Koch P, M\"{u}ller B and Rafelski J 1986
  {\it Phys. Rep.} {\bf 142} 167
\bibitem{Mar} Mar\'{\i}n A {\it et al}. [CERES Collaboration] 2004 
  {\it J. Phys.} G {\bf 30} S709 ({\it Preprint} nucl-ex/0406007)
\bibitem{PPB04} Eidelman S {\it et al}. [Particle Data Group] 2004 
  {\it Phys. Lett.} B {\bf 592} 1
\bibitem{Jovan} Milo\v{s}evi\'c J [CERES Collaboration] 2005 {\it
    Proc. Quark Matter, Nucl. Phys.} {\bf A} in print, ({\it Preprint}
  nucl-ex/0510057); Milo\v{s}evi\'c J 2006 (Heidelberg University) {\it
    Doctoral Thesis} published
\bibitem{Wilrid} Ludolphs W 2006 (Heidelberg University) {\it Doctoral
    Thesis}, published
\bibitem{Dinh99} Dinh P M, Borghini N and Ollitrault J-Y 2000
  {\it Phys. Lett.} {\bf B477} 51
\bibitem{Kolb01} Kolb P F, Huovinen P, Heinz U W and Heiselberg H 2001
  {\it Phys. Lett.} {\bf B500} 232
\bibitem{Pasi05} Huovinen P 2005 Private Communication
\bibitem{Kolb99} Kolb P F, Sollfrank J and Heinz U W 1999
  {\it Phys. Lett.} B {\bf 459} 667
\bibitem{Stef05} Stefanek G 2005 [NA49 Collaboration] {\it Proc. Quark
    Matter, Nucl. Phys.} {\bf A} in print; ({\it Preprint} nucl-ex/0510067)
\bibitem{Oldenburg05} Oldenburg M [STAR Collaboration] 2005
  {\it J. Phys.} G {\bf 31} S437
\bibitem{Cso96} Cs\H{o}rg\H{o} T and L\H{o}rstad B 1996 {\it Phys. Rev.} C
  {\bf 54} 1390
\bibitem{Csa04} Csan\'{a}d M, Cs\H{o}rg\H{o} T and L\H{o}rstad B 2004
  {\it Nucl. Phys.} A {\bf 742} 80
\bibitem{Cso03} Cs\H{o}rg\H{o} T {\it et al}. 2003 {\it Phys. Rev.} C {\bf
    67} 034904
\bibitem{Taran05} Taranenko A [PHENIX Collaboration] 2005 
  ({\it Preprint} nucl-ex/0506019)
\bibitem{Radom} Radomski S 2006 (GSI) {\it Doctoral Thesis} preliminary
\bibitem{NA49} H\"{o}hne C [NA49 Collaboration] 1999 {\it Nucl. Phys.} A
  {\bf 661} 485c
\bibitem{NA49_nuclexp} Mischke A [NA49 Collaboration] 2002 
  ({\it Preprint} nucl-ex/0209002)
\bibitem{NA57} Antinori F {\it et al}. [NA57 Collaboration] 2005
  {\it J. Phys.} G {\bf 31} 1345
\bibitem{PRLMarin} Mar\'{\i}n A {\it et al}. [CERES Collaboration] 2006
  {\it Phys. Rev. Lett.} {\bf 96} 152301 ({\it Preprint} nucl-ex/0512007)
\bibitem{Misko} M\'{\i}skowiec D [CERES Collaboration] 2005 {\it Proc. Quark
    Matter Nucl. Phys.} {\bf A} in print, ({\it Preprint} nucl-ex/0511010)
\bibitem{Sergej} Yurevich S 2006 (Heidelberg University) {\it Doctoral
    Thesis}, published 
\bibitem{Sako} Sako H [CERES Collaboration] 2000 {\it Technical Report} {\bf 03-25}
\bibitem{Rap} Rapp R and Wambach J 2000 {\it Adv. Nucl. Phys.} {\bf 25} 1;
   Rapp R 2005 Private Communication; Braaten E, Pisarski R D and Yuan T-C 1990
   {\it Phys. Rev. Let.} {\bf 64} 2242
   \bibitem{Ror} R\"{o}hrich D 2001 {\it J. Phys.} G {\bf 27} 355
\end{thebibliography}
\end{document}